\documentclass[doublecol]{epl2} 
\usepackage{graphicx}
\usepackage{amsmath,amssymb}
\usepackage{dcolumn}
\usepackage{bm}
\usepackage[latin1]{inputenc}
\usepackage{color}
\newcommand{\rn}[1]{\romannumeral #1}
\newcommand{\RN}[1]{\expandafter\@slowromancap\romannumeral #1@}
\newcommand{\CG}{\mathcal{G}}
\newcommand{\CV}{\mathcal{V}}
\newcommand{\CE}{\mathcal{E}}
\usepackage{amsthm}

\theoremstyle{definition}

\newcommand{\mf}[1]{\mathbf{#1}}

\title{Uncovering hidden flows in physical networks}
\shorttitle{Title} 

\author{Chengwei Wang \inst{1} \and Celso Grebogi\inst{1} \and Murilo S. Baptista\inst{1}}

\institute{                    
  \inst{1} Institute for Complex Systems and Mathematical Biology, University of Aberdeen, King's College, AB24 3UE Aberdeen, United Kingdom.
}

\abstract{
Understanding the interactions among nodes in a complex network is of great importance, since they disclose  how these nodes are cooperatively supporting the functioning of the network.
Scientists have developed numerous methods to uncover the underlying adjacent physical connectivity based on measurements of functional quantities of the nodes states. Often, the physical connectivity, the adjacency matrix, is available. Yet, little is known about how this adjacent connectivity impacts on the ``hidden" flows being exchanged between any two arbitrary nodes, after travelling longer non-adjacent paths.
In this Letter, we show that hidden physical flows in conservative flow networks, a quantity that is usually inaccessible to measurements, can be determined by the  interchange of physical flows between any pair of adjacent nodes.
Our approach applies to steady or dynamic state of either linear or non-linear  complex networks  that can be modelled by  conservative flow networks, such as gas supply networks, water supply networks and power grids.
}

\begin{document}

\maketitle

\section{Introduction}
Research on complex networks \cite{erdds1959random, erd6s1960evolution, watts2004six, PhysRevLett.85.4633, PhysRevEwang, wang2015one, bio.network.1, social.network.1, sensor.network.1, computer.network.2, me.models.morden.powergrid, murilo.kuramoto.powergrid, wang2016data, wang2009abnormal,  helbing2004physics, molkenthin2014networks, watts1999small} and  their applications to real world problems  \cite{wang2016control, me.models.morden.powergrid} have been attracting the attention of many scientists . 
To understand large-scale behaviour of complex networks, it is imperative to calculate the amount of physical flow going  from one node to another one, a quantity that we refer in this work as  ``hidden" flow, since this quantity is usually inaccessible to measurements. 
In this Letter, we avail from the flow tracing method, known in electrical engineering \cite{khan2014power, reta2001electricity, kirschen1997contributions, kirschen1999tracing, bialek1996tracing, jing2005discussion, xie2009analytical, anuar2010observability, malikdetermination, reddytracing}, to calculate the hidden flow between any two nodes, by only requiring information about the adjacent flows between any two connected nodes.
This work  provides a rigorous way to  calculate hidden flows, which in turn enables one to gauge the non-adjacent interactions among nodes in a network,  for networks whose non-adjacent nodes are far apart. 
The applicability of the method is enormous since flow networks can be used as simple models of flow behaviour to many complex networks, such as transportation networks, water supply networks and power grids.
We extend the method to provide an immediate picture of how nodes interact non-adjacently in non-linear networks by constructing linear equivalent models to these networks.

{\bf Flow networks} describe a system that exchanges physical flows. 
Physical flows are usually recognised as the transference of a physical entity (such as the electric charge, a liquid, a solid, a gas volume, cars, airplanes, air, etc) from one node to another in a giving unit of time. 
But they can also be, in a more general sense, probabilities or the information rate (in bits/s). 
In a flow network, there are source nodes that input physical flows (a generator in a power-grid, for example) and sink nodes from which the physical flows leave the network (a consumer in a power-grid, for example).
Flow networks can have several configurations, and for each configuration there are several scientific challenges. 
This work deals with flow networks that are conservative (i.e., total inflow arriving in a node is equal to total outflow leaving it) and whose rule of flow exchange is linear, such as is the case of a direct current electric network. 
Moreover, the edges carrying the flows are uncapacitated, allowing any arbitrary flow intensity.
A remarkable challenge in the area of flow networks is to trace the flow between two non-adjacent nodes (or edges). 
In lieu of studying flows provided by adjacent connections, tracing methods enable one to calculate the amount of flow exchanged from one node (or edge) to another node (or edge), after travelling through several different paths in the network, a quantity being referred in this work as the ``hidden" flow.  
This computationally doable complex task in small flow networks becomes impractical in larger complex flow networks. 
The present work reduces this complicated tracing mathematical process into a trivial manipulation of the so called extended incidence matrix $\mf{K}$ that can be easily calculated from information on the flows along the edges.
We then demonstrate that the hidden flows between any arbitrary pair of nodes can be calculated by our result condensed in Eq. (\ref{master-result}). 
This result, rigorously derived for directed flow networks (preferential direction of flows) and to networks without closed looping flows (where flows circle around a closed path loop) was also extended to the treatment of networks whose flows are undirected and networks that present closed loops. 
Finally, we also show how to extend this result to understand the non-adjacent interactions between any pair of nodes in more general dynamical networks, such as phase oscillator networks, whose behaviour can be well represented by a conservative flow network.

\section{Flow Networks}

\begin{figure}[htb]
\centering
\includegraphics[width=0.8\linewidth]{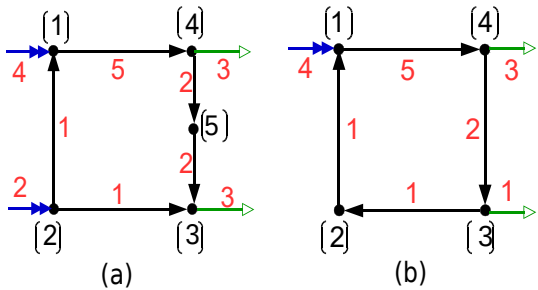}
\caption{(colour online) (a) A flow network without loop flow. (b) A flow network with loop flow.  The black numbers in square brackets are labels of nodes, the red numbers are adjacent flows, the blue lines with double filled arrows are flow sources, the green lines with unfilled arrows are flow sinks, and the black lines with single filled arrows are directed adjacent flows between nodes.}
\label{fig:3.flow}
\end{figure}

A flow network is a digraph, $\CG(\CV,\CE)$,  where $\CV$ and $\CE$ are the sets of nodes and edges, respectively.
A flow network normally contains three types of nodes: (\rn{1}) the source node [e.g., node 1 or 2 in Fig.~\ref{fig:3.flow} (a)], which has a source injecting flow into the network; (\rn{2}) the sink node [e.g., node 3 or 4 in Fig.~\ref{fig:3.flow} (a)], which has a sink taking flow away from the network; (\rn{3}) the junction node [e.g., node 5 in Fig.~\ref{fig:3.flow} (a)], which distributes the flow.
We define $f_{ij}$ to be the \textit{adjacent flow}, or simply the \textit{flow} which is the measurable flow coming from nodes $i$ to $j$ through edge $\{i,j\}\in \CE$.
$f_{ij}=0$ if nodes $i$ and $j$ are not physically connected.
We begin our analysis with the conservative flow networks \cite{ahuja1993network} satisfying: (\rn{1}) $f_{ij}=-f_{ji}$; (\rn{2}) $\sum_{j \in \mathcal{V}} f_{ij}=0$, where node $i$ is a junction node; (\rn{3}) there is no loop flow representing a closed path in a flow network, where a loop flow is shown in Fig.~\ref{fig:3.flow} (b); (\rn{4}) every node must be connected to at least one other node in the network.
A path in a digraph $\CG$ from node $i$ to node $j$, $P(i,j)=i~\{i,i'\}~i'~\{i',i''\}\cdots\{j',j\}~j$, is an alternating sequence of distinct nodes and edges in which the directions of all edges must coincide with their original directions in $\CG$.
The \textit{hidden flow}, $f_{i\rightarrow j}$, is defined to be the summation of the flows going from node $i$ to $j$ through all possible paths from node $i$ to $j$. 

Normally, we can measure or calculate the adjacent flows in a flow network, but it is not easy to obtain the hidden flows, a quantity typically not accessible through measurements.
We find the calculation of hidden flows based on the information of adjacent flows, in a conservative flow network, by the ``flow tracing" method.

Define the \textit{node-net exchanging flow} at node $i$ by 
\begin{equation}
\label{injecting flow}
f_i=\sum_{j=1}^N f_{ij}.
\end{equation}
If node $i$ is a source node, we have $f_i>0$;  we denote $f_i$ by $f_i^s$ as the amount of the \textit{source flow} being injected into the network from a source at node $i$. 
We set $f_i^s=0$ if node $i$ is a sink node or a junction node.
If node $i$ is a sink node we have $f_i<0$;  we denote $f_i^t=-f_i>0$ to indicate the amount of the \textit{sink flow} leaving the network from the sink at node $i$. 
We set $f_i^t=0$ if node $i$ is a source node or a junction node.

Assume there is a positive flow from node $i$ to node $j$, denoted by $f_{ij}>0$.
We use $f_{ij}^{out}$ to indicate $f_{ij}$ as an \textit{outflow} from node $i$ arriving at node $j$, and $f_{ij}^{in}$ to represent $f_{ij}$ as an \textit{inflow} at node $j$ coming from node $i$.
Thus, $f_{ij}=f_{ij}^{out}=f_{ij}^{in}>0$.
$f_{ij}$ can be positive, negative or zero in a flow network.
However, we restrict any outflow or inflow at a node to be a non-negative number.
This means that, if $f_{ij}<0$, we force $f_{ij}^{out}$ and $f_{ij}^{in}$ to be zeros.
Analogously, $f_{ij}<0$ means $f_{ji}>0$, we have $f_{ji}^{out}>0$ to denote the outflow from node $j$ to node $i$ and $f_{ji}^{in}>0$ to be the inflow at node $i$ from node $j$.

Define the \textit{total inflow} at node $i$ by 
\begin{equation}
\label{total inflow}
f_i^{in}=f_i^s+\sum_{f_{ji}>0} f_{ji}=f_i^s+\sum_{j=1}^N f_{ji}^{in},
\end{equation}
and the \textit{total outflow} at node $i$ by 
\begin{equation}
\label{total out}
f_i^{out}=f_i^t+\sum_{f_{ij>0}} f_{ij}=f_i^t+\sum_{j=1}^N f_{ij}^{out}.
\end{equation}
In a conservative flow network, the total inflow of a node is equal to its total outflow, i.e., $f_i^{out}=f_i^{in}$.
We assume $f_i^{out}=f_i^{in}> 0$, $\forall i$, meaning that each node in a flow network must exchange flow with other nodes, i.e., no node is isolated.

\section{Flow tracing by proportional sharing principle} 
\begin{figure}[htb]
\centering
\includegraphics[width=0.5\linewidth]{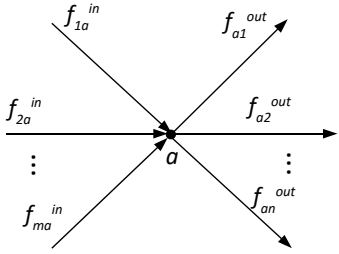}
\caption[Inflow and outflow]{A node $a$ with $m$ inflows and $n$ outflows.}
\label{fig.3.proportion node}
\end{figure}

The \textit{proportional sharing principle (PSP)} \cite{liu2000proof, jing2005discussion} states that  for an arbitrary node, $a$, with $m$ inflows and $n$ outflows (Fig.~\ref{fig.3.proportion node}) in a conservative flow network, (\rn{1}) the outflow on each outflow edge is proportionally fed by all inflows,
and (\rn{2}) by assuming  that node $i$ injects a flow $f_{ia}^{in}$ to node $a$, and node $j$ takes a flow $f_{aj}^{out}$ out of node $a$, we have that the \textit{node-to-node hidden flow}  from node $i$ to  node $j$ via node $a$ is calculated by
\begin{equation}
\label{eq.3.proportional sharing principle down}
f_{i\rightarrow j}=f_{ia}^{in}\frac{f_{aj}^{out}}{f^{out}_a},
\end{equation}
or by
\begin{equation}
\label{eq.3.proportional sharing principle up}
f_{i\rightarrow j}= f_{aj}^{out}\frac{f_{ia}^{in}}{f_a^{in}}.
\end{equation}

Equations~(\ref{eq.3.proportional sharing principle down}) and (\ref{eq.3.proportional sharing principle up}) result in the same value of $f_{i\rightarrow j}$, since $f_a^{out}= f_a^{in}$.
Equation~(\ref{eq.3.proportional sharing principle down}) represents the \textit{downstream flow tracing} method, where we start tracing the hidden flow from a source node $i$ to a sink node $j$, by using the percentage, $f_{aj}^{out}/f_{a}^{out}$, to indicate the percentage of $f_{ia}^{in}$ that goes to $j$.
Equation~(\ref{eq.3.proportional sharing principle up}) denotes the \textit{upstream flow tracing} method, where we trace the flow from a sink node $j$ to a source node $i$, by knowing the proportion of $f_{aj}^{out}$ is provided by $f_{ia}^{in}$.

The percentage $f_{aj}^{out}/f_{a}^{out}$ in Eq.~(\ref{eq.3.proportional sharing principle down}) and $f_{ia}^{in}/f_{a}^{in}$ in Eq.~(\ref{eq.3.proportional sharing principle up}) are related to the flows on edges.
They are similar to the probability of jumping from a node  to  one of its neighbours in a biased random walk process \cite{sinatra2011maximal, gomez2008entropy, lambiotte2011flow}, where a similar percentage is related to the weight of edges.

We only deal with the downstream flow tracing in the Letter and explain the upstream flow tracing  in the Supplementary Material \cite{sup}.

Define the \textit{downstream coefficient} at node $a$ for the outflow $f_{aj}^{out}$ by
\begin{equation}
\kappa^{d}_{aj}=\frac{f_{aj}^{out}}{f_a^{out}},
\end{equation}
to indicate the proportion of the outflow at edge $\{a,j\}$ to the total outflow at node $a$.
Define the \textit{upstream coefficient} at node $a$ for the inflow $f_{ia}^{in}$ by
\begin{equation}
\kappa^{u}_{ai}=\frac{f_{ia}^{in}}{f_a^{in}},
\end{equation}
denoting the proportion of the inflow at edge $\{i,a\}$ to the total inflow at node $a$.
Then the calculation of $f_{i\rightarrow j}$ can be simply expressed by $f_{i\rightarrow j}=f_{ia}^{in} \kappa_{aj}^d$ or $f_{i \rightarrow j}=f_{aj}^{out} \kappa_{ai}^u $.

Define the \textit{sink proportion} and \textit{source proportion} at node $a$ by
\begin{equation}
\label{sink and source proportion}
\iota^t_{a}=\frac{f_a^t}{f_a^{out}} \text{ and } \iota^s_{a}= \frac{f_a^s}{f_a^{in}},
\end{equation}
respectively, where the sink proportion, $\iota^t_{a}$, indicates the proportion of the sink flow to the total outflow at node $a$, and the source proportion, $\iota^s_{a}$, indicates the proportion of the source flow to the total inflow at node $a$.
By defining the sink proportion and source proportion, we are now able to calculate the \textit{source-to-sink hidden flow} from a source at node $i$ to a sink at node $j$ denoted by $f_{si\rightarrow tj}$.
From Eq.~(\ref{total inflow}), we know that $f_i^s$ is a part of $f_i^{in}$, where $f_i^s$ is the source flow at node $i$.
From Eq.~(\ref{sink and source proportion}), we know the proportion of $f_i^s$ to $f_i^{in}$.
According to the PSP, we can then calculate the source-to-sink hidden flow by $f_{si\rightarrow tj}=\frac{f_i^s}{f_i^{in}} f_{i \rightarrow j} \frac{f_{j}^t}{f_{j}^{out}}=\iota_i^s f_{i \rightarrow j} \iota_{j4W8QF9-3DS84E}^t$.

It is possible to trace (calculate) the hidden flows from any arbitrary pair of nodes in a flow network using either the downstream or the upstream approach. 
However, all the paths connecting a pair of nodes must be considered. 
In particular, the hidden flow from two adjacent nodes will include the flow exchanged along the adjacent connection and all the flows travelling along other longer paths connecting these two adjacent nodes. 
Suppose one wants to calculate the hidden flow $f_{i \rightarrow j}$ from two non-adjacent nodes $i$ and $j$, and there are two possible paths, $P_1(i, j) = i \{i, k\} \{k, j\} j$ and  $P_2(i,j) = i \{i, l\} \{l, g\} \{g, j\} j$, $P_1$ with length 2 and $P_2$ with length 3. 
Each path produces a hidden flow,  $f^{(1)}_{i \rightarrow j}$ and $f^{(2)}_{i \rightarrow j}$, respectively. 
The total hidden flow from $i$ to $j$ is thus calculated using that $f_{i \rightarrow j} = f^{(1)}_{i \rightarrow j} + f^{(2)}_{i \rightarrow j}$, where $f^{(1)}_{i \rightarrow j} = f^{in}_i \kappa_{ik}^d \kappa_{kj}^d$ and   $f^{(2)}_{i \rightarrow j} = f^{in}_i \kappa_{il}^d \kappa_{lg}^d \kappa_{gj}^d$. 
This process is feasible when dealing with small flow networks, as illustrated in the Supplementary Material \cite{sup}, where we show how to trace hidden electric current flows in a direct current (DC) electric network. 
But it becomes impractical when dealing with large networks, for which the number of paths carrying flows can grow exponentially fast with the size of the network. To circumvent this challenging calculation, the use of the extended incidence matrix, $\mf{K}$, proposed in Refs. \cite{xie2009analytical, anuar2010observability, malikdetermination}, is taken forward.

\section{Flow tracing by extended incidence matrix}
The \textit{downstream extended incidence matrix}, $\mf{K}$, in a flow network with $N$ nodes is an $N\times N$ dimensional matrix, defined by
\begin{equation}
\label{eq.3.downstream extended incidence matrix}
K_{ij}=
\begin{cases}
-f_{ji}^{in}/f_j^{out} & ~\text{if}~i\neq j,~\text{and}~f_{ji}>0,\\
1 & ~\text{if}~i=j,\\
0 & ~\text{else}.
\end{cases}
\end{equation}

Transform Eq~(\ref{total inflow}) to
$
f_i^{in}-\sum_{j=1}^{N} f_{ji}^{in}/f_j^{out} \cdot f_j^{out}=f_{i}^s.
$
Considering $f_i^{in}= f_i^{out}$, we have
\begin{equation}
\label{eq.3.downstream equaiton}
f_i^{out}-\sum_{j=1}^{N} f_{ji}^{in}/f_j^{out} \cdot f_j^{out}=f_{i}^s.
\end{equation}
From Eqs.~(\ref{eq.3.downstream extended incidence matrix}) and (\ref{eq.3.downstream equaiton}), we have
\begin{equation}
\label{eq.3.downstream matrix equation}
\mf{K F^{out}=F^s},
\end{equation}
where $\mf{F^{out}}=[f_{1}^{out},~f_{2}^{out},\cdots, f_{N}^{out}]^T$, and $\mf{F^s}=[f_1^s,~f_2^s,\cdots,f_N^s]^T$.
$\mf{K }$ is an invertible matrix \cite{xie2009analytical, malikdetermination, reddytracing}, thus, $\mf{F^{out}=K^{-1}F^s}$, implying that,
\begin{equation}
\label{eq.3.downstream matrix coefficient 1}
f_{i}^{out}=\sum_{j=1}^N\left[\mf{K^{-1}}\right]_{ij} f_j^s,
\end{equation}
$\left[\mf{K^{-1}}\right]_{ij}$ being an entry ($i^{th}$ row, $j^{th}$ column) of the matrix $\mf{K^{-1}}$.
Equation~(\ref{eq.3.downstream matrix coefficient 1}) indicates that the outflow of node $i$, $f_i^{out}$, is fed by every source $f_j^s$. More specifically, $\mf{K^{-1}}_{ij}$ represents the proportion of the source inflow in the source node $j$ that goes to node $i$.

Let $\mf{C=K^{-1}}$ be the \textit{downstream contribution matrix}.
Considering $\iota_j^s={f_j^s}/{f_j^{in}}$, we have
\begin{equation}
\label{downstream contri}
f_{i}^{out}=\sum_{j=1}^N C_{ij}f_j^{in} \iota_j^s.
\end{equation}
Knowing that the source-to-node hidden flow from source node $j$ to node $i$ is given by $
f_{sj\rightarrow i}=\iota_j^s f_{j\rightarrow i}$, Eq. (\ref{downstream contri}) thus implies that for 
a source node $j$ with $\iota_j^s\neq 0$, $C_{ij}f_j^{in}$ represents the node-to-node hidden flow from node $j$ to node $i$, i.e., $f_{j\rightarrow i}=C_{ij}f_j^{in}$. 
The tracing of flows from source to nodes, previously known in the literature, only applied to source nodes. 
To extend it to any other general situation, including the tracing of flows from and to edges, sinks and junction nodes, we introduce an \textit{equivalence principle}.
We treat any sink or junction node as a hypothetical source node, without altering the original network topology and flows.   
If node $j$ is a sink or junction node with a total inflow $f_{j}^{in}>0$ and $\iota_j^s= 0$, we treat node $j$ as a hypothetical source node with $f_j^{s}=f_{j}^{in}>0$, where the hypothetical source takes the place of all the edges injecting flows into $j$.
By this treatment, we can hypothetically treat node $j$ as a source node with $\iota_j^s=f_j^{s}/f_{j}^{in}=1$, in Eq. (\ref{downstream contri}), such that the node-to-node hidden flow from node $j$ to node $i$ can also be calculated by 

\begin{equation}
\label{master-result}
f_{j\rightarrow i}=C_{ij}f_j^{in}.
\end{equation}

Thus, from our analysis, $C_{ij}=\left[\mf{K^{-1}}\right]_{ij}$ is a \textit{donwstream contribution factor} indicating how much hidden flow goes from node $j$ to $i$, i.e., $f_{j\rightarrow i}=C_{ij} f_j^{in}$ for any pair of nodes.
%
%

Now, we show how non-adjacent hidden flows can be traced in conservative flow networks. Notice for networks whose non-adjacency nodes are far apart from each other, the hidden flows can gauge how non-adjacent interactions emerge in the studied system.
Let $i$, $j$, $m$, $n$, $p$, $q$ be different nodes in a conservative flow network, where node $i$ has a source, node $j$ has a sink, nodes $m$, $n$ are connected by edge $\{m,n\}$ with $f_{mn}>0$, and nodes $p$, $q$ are connected by edge $\{p,q\}$ with $f_{pq}>0$. 
The non-adjacent interaction includes:
(\rn{1}) the node-to-node hidden flow from node $i$ to $j$ is
$
f_{i\rightarrow j}=C_{ji}f_i^{in};
$

(\rn{2}) the source-to-node hidden flow from source node $i$ to node $j$ is
$
f_{si\rightarrow j}=\iota_i^s f_{i\rightarrow j};
$
(\rn{3}) the node-to-sink hidden flow from source node $i$ to sink node $j$ is
$
f_{i\rightarrow tj}=f_{i\rightarrow j} \iota_j^t;
$
(\rn{4}) the source-to-sink hidden flow from node $i$ to $j$ is
$
f_{si\rightarrow tj}=\iota_i^s f_{i\rightarrow j}\iota_j^t;
$
(\rn{5}) the node-to-edge hidden flow from node $i$ to edge $\{m,n\}$ is
$
f_{i\rightarrow \{m,n\}}=f_{i\rightarrow m}\cdot \kappa^{d}_{mn}
$;
(\rn{6}) the edge-to-node hidden flow from edge $\{m,n\}$ to node $j$ is
$
f_{\{m,n\}\rightarrow j}=\kappa^{u}_{nm} \cdot f_{n\rightarrow j};
$
and (\rn{7}) the edge-to-edge hidden flow from edge $\{p,q\}$ to $\{m,n\}$ is
$
f_{\{p,q\}\rightarrow \{m,n\}}=\kappa^{u}_{qp} \cdot f_{q\rightarrow m} \cdot \kappa^{d}_{mn}$.

To illustrate the calculation of these hidden flows, as well as the calculation of the matrices involved in it, in the Supplementary Material \cite{sup} we trace the flows in an electric network using our downstream extended incidence matrix approach.

\section{Extension to flow networks with closed loops and with undirected flows}\label{extension}

{\bf Loops:} If the closed loop (or loops) is inside a larger network, one needs first to identify the existence of a loop. 
A closed loop at the node $i$ with a length $P$ exists in a network if $[A^P]_{ii}>0$, where $[A^P]_{ii}$ represents the term $ii$ in the power to $P$ of the adjacency matrix of the network. 
The source node of the loop is any node receiving input flow, and the sink node is the one containing an edge with an outflow, and whose path length connecting it to the source node is the longest.

We consider a network with 4 nodes, with a loop flow as in Fig. \ref{fig:3.flow}(b).
Let us call it network $N$. 
Denote the input flow as $f_1^s(N)$, the output flows as $f_4^t(N)$ and $f_3^t(N)$, and the adjacent flows as $f_{14}(N)$, $f_{43}(N)$, $f_{32}(N)$, and $f_{21}(N)$. 
A loop in a flow network is broken down into subnetworks in which the flows are directed.
Merging the flows of all subnetworks must preserve edge, source and sink flows of the original network $N$. 
In Fig. \ref{fig:3.flow}(b), the loop is formed by $1\{1,4 \} \{4,3\} \{3,2\} \{2,1\}1$. 
To break-up the loop, one firstly choose a source and a sink node, where flows enter and leave the closed loop, respectively. 
Node 1 is the only source node. 
The sink node to be chosen must be the one whose length of a direct path connecting it to the source node is the longest one. 
We choose node 3 as the sink node.
Then, one needs to determine all the directed paths connecting the source node (node 1) and to the sink node (node 3), and all the directed paths connecting the sink to the source nodes. 
Among all paths, one takes only the paths that have the same flow directions as the original network $N$.
These directed paths form the subnetworks whose net flow represents the original network flow and from which the hidden flows are calculated. 

\begin{figure}[htb]
\centering
\includegraphics[width=7cm]{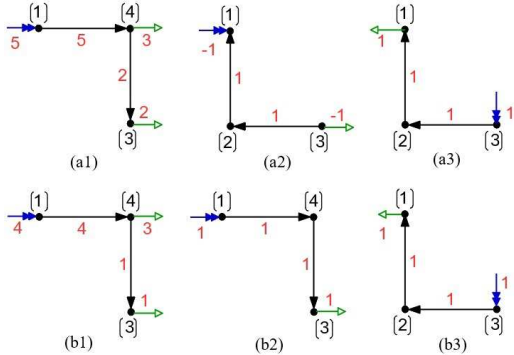}
\caption{(colour online) Illustrations of two approaches to break-up a flow network with a closed loop flow into smaller subnetworks with only directed flows.}
\label{fig.loops}
\end{figure}

We show, in Fig.~\ref{fig.loops}, the subnetworks of the network in Fig. \ref{fig:3.flow}(b).
Panel (a1) represents a directed path and its flows from node 1 to node 3.
Panels (a2) and (a3), with the same directed path subnetwork, show the directed paths connecting nodes 3  to 1 . 
Notice that a negative source and sink, in nodes 1 and 3,  respectively, in panel (a2), is equivalent to a positive sink and source nodes, respectively, as represented in panel (a3).  
In panels (b1)-(b3), we show another practical way to determine the break up of the network with a closed loop. 
Once a loop, and a source and a sink nodes, are identified, we remove it from the network. 
Panel (b1) is the subnetwork after the loop removal.
The closed loop is formed by merging the flows represented in panels (b2) and (b3), and it has a constant flow of 1 unit. 
One restores the original network by adding the subnetworks in panel (a1) and (a3), or by adding the subnetworks in panels (b1), (b2), and (b3). 
Calculating hidden flows of the original network needs to take into consideration of hidden flows in all subnetworks. 
One subnetwork [panel (a1)], let us call it $N1$, is formed by the nodes 1, 3, and 4.
Node 2 is absent and, therefore, to preserve edge flows one is required to make $f_1^s(N1)=f_1^s+f_{21}(N)$ and 
$f_3^t(N1)=f_3^t+f_{32}(N)$. 
From this network, $f_{1 \rightarrow 4}=5$, $f_{s1 \rightarrow t4}=3$, $f_{s1 \rightarrow t3}=2$.  
The other network [panel (a2)], let us call it $N2$, is formed by the nodes  1, 2, and 3, so node 4 is now absent and therefore, to preserve edge flows we are required to make $f_1^s(N2)=f_1^s+f_{41}(N)=f_1^s-f_{14}(N)$ and $f_3^t(N2)=f_3^t+f_{34}(N)=f_3^t-f_{43}(N)$. 
These equations lead to  $f_1^s(N2)<0$ and $f_3^t(N2)<0$, whose flows are indicated in panel (a2). 
The hidden flow from node 2 and 4 is zero, since no subnetworks contribute to a hidden flow from node 2 to 4.

{\bf Undirected flow networks:   Similarly, our method can also be applied to an undirected  flow network if the network can be split into two independent unidirectional networks.  
For example, under the assumption that all traffic roads are bidirectional, we can separate the transportation network of a city into two networks. 
One network includes all the left-hand roads and the other one contains all the right-hand roads. 
Thus, both separated networks become unidirectional networks.}

\section{Non-adjacent interaction in non-linear networks}
Next, we extend our tracing hidden flow approach to study non-linear systems by constructing linear model analogous to the non-linear networks. 
Let the equation
\begin{equation}
\label{eq.3.dynamic system}
\dot{x_i}=S(x_i)-\sum_{j=1}^N L_{ij}\cdot H(x_i,x_j)
\end{equation} 
indicate a dynamic scheme describing the behaviour of $N$ coupled nodes, where $x_i$ is the dynamical variable of each node, $S(x_i)$ is the isolated dynamic function, $L_{ij}$ is the element of the Laplacian matrix, and $H(x_i,x_j)$ is an arbitrary coupled dynamic function.
We treat the system as a flow network by interpreting $f_i(t)=S(x_i)-\dot{x}_i$  as the node-net exchanging flow at node $i$.
The value and sign of $f_i(t)$ may change over time.
If $f_i(t)>0$ (or $f_i(t)<0$), we treat node $i$ as a source (or sink) node at time $t$ and the source (or sink) flow is $f_i^s(t)=f_i(t)$ (or $f_i^t(t)=-f_i(t)$).
If $f_i(t)=0$, we treat node $i$ as a junction node at time $t$.
Let $f_{ij}(t)=L_{ij} H(x_i,x_j)$ be the adjacent flow from node $i$ to node $j$.
If $f_{ij}(t)>0$, we have $f_{ij}^{out}(t)>0$ as the outflow from node $i$ and $f_{ij}^{in}(t)>0$ as the inflow at node $j$ at time $t$.
If $f_{ij}(t)<0$, we have $f_{ji}^{out}(t)>0$ as the outflow from node $j$ and $f_{ji}^{in}(t)>0$ as the inflow at node $i$ at time $t$.
By doing this interpretation, we are constructing an equivalent linear conservative flow network that  behaves in the same way as the non-linear network described by Eq.~(\ref{eq.3.dynamic system}).
This enables us to calculate the non-adjacent interactions in the equivalent linear flow network which informs us about the non-adjacent interactions in the original non-linear network.

We consider a revised Kuramoto model \cite{kuramoto1975, kuramoto1984, kuramoto1987} as an example, which is given by
\begin{equation}
\label{eq.4 Kuramoto model}
\dot{\theta}_i=\omega_i-K\sum_{j=1}^{N}L_{ij}\sin(\theta_i-\theta_j),
\end{equation}
where $K$ is the coupling strength, $L_{ij}$ is the entry of the Laplacian matrix, $\theta_i$ and $\omega_i$ indicate the phase angle and natural frequency in a rotating frame, respectively.
In this rotating frame, $\dot{\theta}_i=\dot{\theta}_j=0,~\forall i\neq j$, when the oscillators emerge into \textit{frequency synchronisation (FS)} for a large enough $K$ \cite{kuramoto.onset}.
In the FS state, all the node-net exchanging flows $f_i=\omega_i-\dot{\theta}_i=\omega_i$ and all the adjacent flows $f_{ij}=KL_{ij}\sin(\theta_i-\theta_j)$ are constants, since $\sin(\theta_i-\theta_j)$ are constants.

Let $\alpha_{ij}=|f_{ij}|/\max\{|f_{ij}|:\forall i, j\}$ be a normalised variable in [0,1] indicating the adjacent interaction strength between oscillator $i$ and $j$, where  $\max\{|f_{ij}|:\forall i, j\}$ is the maximum of all absolute values of adjacent flows.
Since $f_{ij}=-f_{ji}$, we have $\alpha_{ji}=\alpha_{ij}$.
Every hidden flow is traced by considering that flows are directed.
This implies that all the calculated hidden flows   are non-negative and at least one of $f_{i\rightarrow j}$ and $f_{j\rightarrow i}$ is 0.
We let $\beta_{ij}=\beta_{ji}=\max\{f_{i\rightarrow j},f_{j\rightarrow i}\}/\max\{f_{i\rightarrow j}:\forall i, j\}$ be the non-adjacent interaction strength between oscillator $i$ and $j$, 
where $\max\{f_{i\rightarrow j},f_{j\rightarrow i}\}$ is the non-zero one between $f_{i\rightarrow j}$ and $f_{j\rightarrow i}$, and $\max\{f_{i\rightarrow j}:\forall i, j\}$ is the maximum of all hidden flows.
This definition of the non-adjacent interaction strength allows us to compare $\alpha_{ij}$ and $\beta_{ij}$ for the same pair of nodes in a network.

\begin{figure}[htb]
\centering
\includegraphics[width=\linewidth,height=4.6cm]{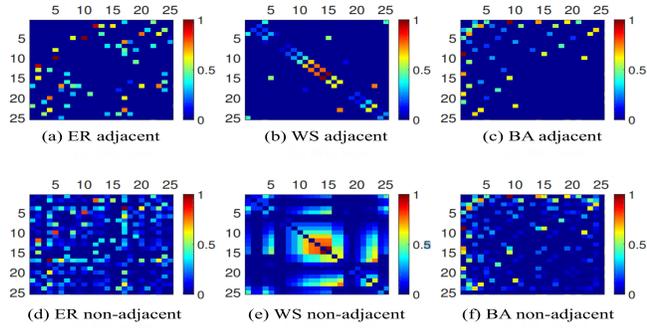}
\caption{(colour online) Comparison of adjacent interactions and non-adjacent interactions in different networks described by the Kuramoto model after the occurrence of frequency synchronisation. (a), (b) and (c) demonstrate the adjacent interactions in ER network, WS network and BA network, respectively, compared with the non-adjacent interactions shown in (d), (e) and (f) for these networks. The numbers on axes are labels of nodes. The colours on maps indicate the interacting strength between nodes.}
\label{fig.comparison of local and non-local}
\end{figure}
We construct three types of networks with 25 nodes, namely the Erd\"{o}s-R\'{e}nyi  (ER)  \cite{erdds1959random, gilbert1959random}, Watts-Strogatz (WS)  \cite{watts1998collective} and Barab\'{a}si-Albert (BA) models \cite{barabasi1999emergence}.
The dynamic behaviour of the nodes in these networks follows Eq.~(\ref{eq.4 Kuramoto model}).
Figure~\ref{fig.comparison of local and non-local} shows the comparison of the adjacent interactions and the non-adjacent interactions when the oscillators emerge into FS with a large enough $K$.
Figures~\ref{fig.comparison of local and non-local} (a), (b) and (c) show the adjacent interaction strengths, $\alpha_{ij}$, for ER, WS and BA networks, respectively.
Figures~\ref{fig.comparison of local and non-local} (d), (e) and (f) demonstrate the non-adjacent interaction strengths, $\beta_{ij}$, for ER, WS and BA networks, respectively.
Figure~\ref{fig.comparison of local and non-local} (d) exposes some hidden interactions that Fig.~\ref{fig.comparison of local and non-local} (a) does not show to exist in an ER network.
By comparing Figs.~\ref{fig.comparison of local and non-local} (b) and (e), we see that a randomly rewired edge in a WS network not only produces interaction between the two adjacent nodes connected by this edge, but also creates functional clusters among nodes close to the two adjacent nodes.
So, complex systems can in fact be better connected than previously thought.
We constructed the BA network by assigning smaller labels to nodes with larger degrees.
Both Figs.~\ref{fig.comparison of local and non-local} (c) and (f) illustrate the strong interactions among the nodes with large degrees (small labels).
Figure~\ref{fig.comparison of local and non-local} (c) shows that the interactions between unconnected nodes with small degrees (large labels) are weak or inexistent, though, such interactions are revealed in Fig.~\ref{fig.comparison of local and non-local} (f). 
Through this comparison, we understand that two nodes in a network may strongly interact with each other even if they are not connected by an edge.

\begin{figure}[htb]
\includegraphics[width=\linewidth, height=4.6cm]{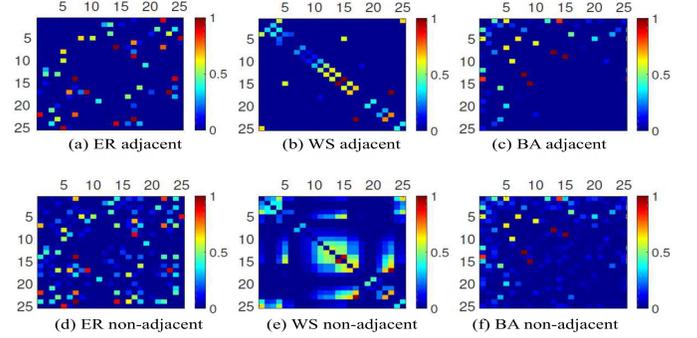}
\caption{(colour online) Comparison of adjacency interactions and non-adjacency interactions in different types of networks described by the Kuramoto model when frequency synchronisation is inexistent. (a), (b) and (c) demonstrate the adjacent interaction strength in ER network, WS network and BA network, respectively.  (d), (e) and (f) show the non-adjacent interaction strength for these networks. The numbers on axes are labels of nodes. The colour on map indicates the interacting strength between nodes.}
\label{fig.comparison of local and global before FS}
\end{figure}

Figure~\ref{fig.comparison of local and global before FS} shows the simulations results of the adjacent interaction strength and non-adjacent interaction strength for these networks when FS is not present.
Final results are taken by averaging the results of 100 time-points that are uniformly chosen in the time scale [10,20], i.e., $\alpha_{ij}=\sum_{k}^{100} \alpha_{ij}(t_k)/100$ and $\beta_{ij}=\sum_{k}^{100} \beta_{ij}(t_k)/100$, where $\alpha_{ij}(t_k)$ and $\beta_{ij}(t_k)$ are the values of $\alpha_{ij}$ and $\beta_{ij}$ at the $k^{th}$ time-point. 
The dynamic behaviour of the oscillators in these networks is described by the Kuramoto model by assigning a small coupling strength, such that the oscillators are in an incoherent state.
Comparing the results in Fig.~\ref{fig.comparison of local and global before FS} with that when FS is present, we find that those pairs of nodes which are not interacting through hidden flows when FS is not present, also present no evident non-adjacency interactions when FS is present.
This suggests that the existence of non-adjacent interaction between a pair of nodes strongly depends on the network topological features of the network rather than the coupling strength.

\section{Conclusion}
In this Letter, we introduced the proportional sharing principle and the extended incidence matrix to calculate the hidden flows in flow networks, and further extended this approach to trace the non-adjacent hidden flows in non-linear complex systems which can analogously be represented by linear flow networks. This allows us to understand the non-adjacency interactions among nodes either under a steady state (e.g., when FS is present in the Kuramoto model) or a dynamic state (e.g., when FS is not present in the Kuramoto model) in such a complex system. Our study illustrated that the nodes in a network not only interacts with their neighbours, but can also strongly influence those who are not directly connected to them. By comparing the results of the non-adjacent study for the Kuramoto model when FS is present and that when FS is not present for different topological networks, we concluded that the emergence of non-adjacent interaction between a pair of nodes  strongly depends on the topological features of the networks rather than the coupling strength between nodes.

We have extended our analysis to flow networks that present closed loops and for those that present undirected flows. The solution for these challenging problems is to break the network into subnetworks that only contain directed flows. The method can also be applied to weighted networks, as long as the weighted network can be modelled as a conservative flow network.

This work opens up a new area of research into non-adjacent interactions in complex networks, facilitating and enabling research that aims at unravelling complex behaviour as a function of the network topology.
There is also great potential to link this work to other works in the area of complex networks, such as the link prediction problem \cite{lu2015toward}, and to the study of information and energy transmission in complex networks \cite{chen2016energy, huang2006information, wang2017physical}.  These potential extentions will further widen the applicability of the method in the real world. 
It is worth mentioning that our work assumed at the outset that the adjacency matrix of the system as well as the adjacency physical flows is known \textit{a priori}. 
Therefore, works such as those in Ref. \cite{lu2015toward} predicting the existence of a physical link should be used prior to our method.

\acknowledgments
Chengwei Wang is supported by a studentship funded by the College of Physical Sciences, University of Aberdeen.


\newpage

\onecolumn
Supplementary Material for\\
\hfill
\begin{center}
\textbf{Uncovering hidden flows in physical networks}
\end{center}

\section{Example of Flow Tracing in a DC Network}

\begin{figure}[htb]
\centering
\includegraphics[width=\linewidth]{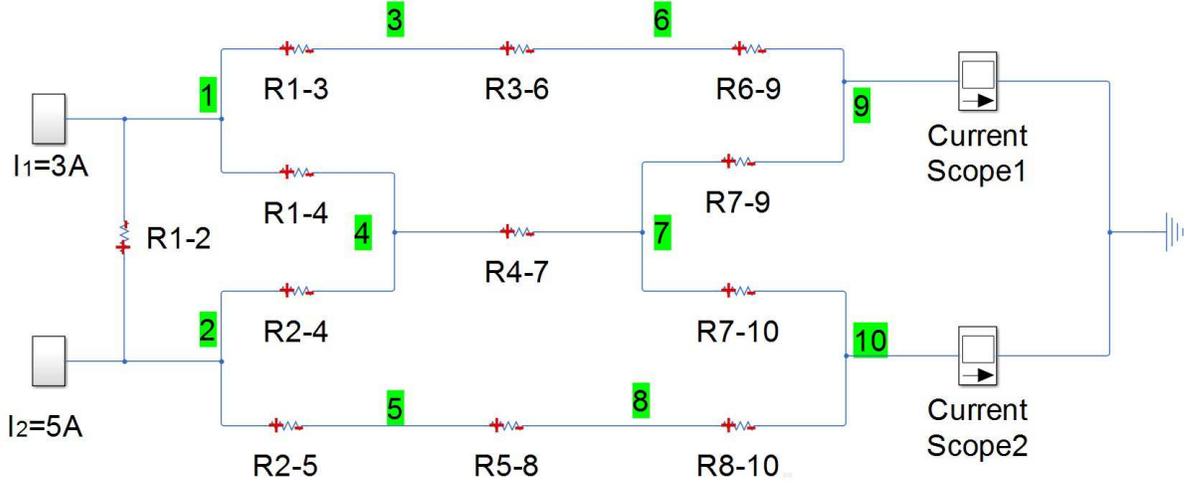}
\caption[The MATLAB/Simulink model for a DC network]{The MATLAB/Simulink model for a DC network with 10 nodes.}
\label{fig.3.DC for simulink}
\end{figure}

\begin{figure}[htb]
\centering
\includegraphics[width=0.5\linewidth]{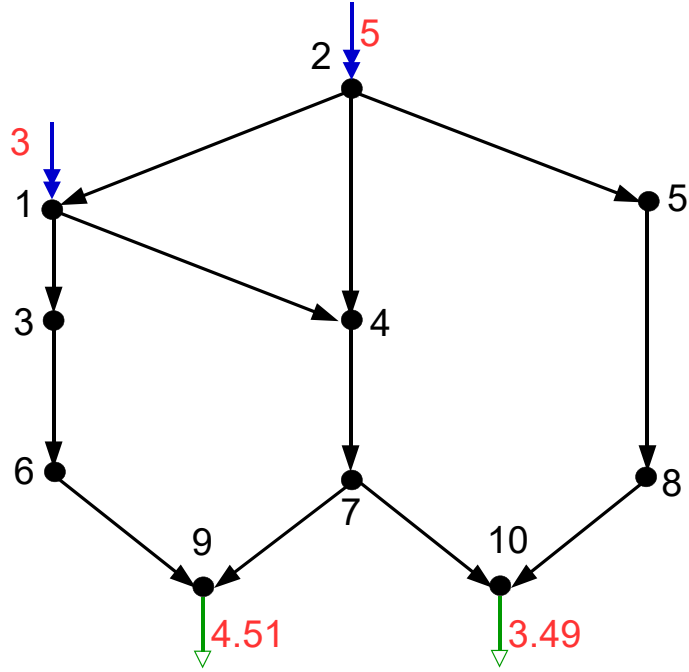}
\caption[The current direction for a DC network]{The current directions in  the DC network shown in Fig.~\ref{fig.3.DC for simulink}.}
\label{fig.3.DC for simulink abstract}
\end{figure}

We build up a MATLAB model to simulate a direct current (DC) network shown in Fig.~\ref{fig.3.DC for simulink} to illustrate the flow tracing process.
The flow quantity $f$ is given by the electric current $I$ in this model.
Nodes 1 and 2 are two nodes with current sources where $I_1^s=3$A and $I_2^s=5$A, respectively.
The resistances of resistors are randomly chosen within the set of integer numbers [1,10], shown in Tab.~\ref{tab.3.resistance of resistors}.
The sink flow leaving from the sink nodes 9 and 10 are measured by the current scopes as $I_9^t=4.51$A and $I_{10}^t=3.49$A.
The current directions are shown in Fig.~\ref{fig.3.DC for simulink abstract}.
Next, we show how to calculate the source-to-sink hidden currents from the current source  $I_1^s$ and $I_2^s$ to the sink $I_9^t$ and $I_{10}^t$ by different methods.

\begin{table}[htb]
\centering
\caption[Resistance of resistors]{Resistances of the resistors in Fig.~\ref{fig.3.DC for simulink}.}
\label{tab.3.resistance of resistors}
\begin{tabular}{|l|c|c|c|c|c|c|}
\hline
Resistor            & $R_{1-2}$  & $R_{1-3}$   &  $R_{1-4}$  & $R_{2-4}$  & $R_{2-5}$  & $R_{3-6}$  \\ \hline
Resistance/$\Omega$ & 7 & 9 & 7 & 4  & 6  & 5  \\ \hline \hline
Resistor            & $R_{4-7}$  & $R_{5-8}$   &  $R_{6-9}$  & $R_{7-9}$  & $R_{7-10}$  & $R_{8-10}$  \\ \hline
Resistance/$\Omega$ & 1 &3 & 2 & 2 & 3 & 8  \\ \hline
\end{tabular}
\end{table}

\subsection{Using the Downstream Flow Tracing Method}
As shown in Fig.~\ref{fig.3.DC for simulink abstract}, there are two paths from node 1 to node 9, which are $P_1(1,9)=1~\{1,3\}~3~\{3,6\}~6~\{6,9\}~9$, and $P_2(1,9)=1~\{1,4\}~4~\{4,7\}~7~\{7,9\}~9$.

Using the downstream flow tracing method, we  calculate the current from node 1 to node 9  through the path $P_1(1,9)$ by
\begin{equation}
I_{1\rightarrow 9}^{(1)}=I_1^{in} \frac{I_{13}^{out}}{I_1^{out}} \frac{I_{36}^{out}}{I_3^{out}} \frac{I_{69}^{out}}{I_6^{out}}=I_1^{in}\kappa_{13}^d \kappa_{36}^d \kappa_{69}^d,
\end{equation}
and through the path $P_2(1,9)$ by
\begin{equation}
I_{1\rightarrow 9}^{(2)}=I_1^{in} \frac{I_{14}^{out}}{I_1^{out}} \frac{I_{47}^{out}}{I_4^{out}} \frac{I_{79}^{out}}{I_7^{out}}=I_1^{in}\kappa_{14}^d \kappa_{47}^d \kappa_{79}^d.
\end{equation}
Thus, the total node-to-node hidden current from node 1 to node 9 is 
\begin{equation}
\label{eq.3.flow calculation node-node}
I_{1\rightarrow 9}=I_{1\rightarrow 9}^{(1)}+I_{1\rightarrow 9}^{(2)}.
\end{equation}
The source-to-sink hidden current is calculated by 
\begin{equation}
\label{eq.3.flow calculation source-sink}
I_{s1\rightarrow t9}=\iota_1^s \cdot I_{1\rightarrow 9}\cdot \iota_9^t.
\end{equation}
By doing this type of calculation, we obtain $I_{s1\rightarrow t9}=2.35$, $I_{s1\rightarrow t10}=0.65$, $I_{s2\rightarrow t9}=2.16$ and $I_{s2\rightarrow t10}=2.84$.
\subsection{Using the Upstream Flow Tracing Method}
Using the upstream flow tracing method, we have 
\begin{equation}
\label{eq.3.flow calculation 1 u}
I_{1\rightarrow 9}^{(1)}=I_9^{out}\frac{I_{69}^{in}}{I_9^{in}}\frac{I_{36}^{in}}{I_6^{in}}\frac{I_{13}^{in}}{I_3^{in}}=I_9^{out} \kappa_{96}^u \kappa_{63}^u \kappa_{31}^u,
\end{equation}
and
\begin{equation}
\label{eq.3.flow calculation 2 u}
I_{1\rightarrow 9}^{(2)}=I_9^{out}\frac{I_{79}^{in}}{I_9^{in}}\frac{I_{47}^{in}}{I_7^{in}}\frac{I_{14}^{in}}{I_4^{in}}=I_9^{out} \kappa_{97}^u \kappa_{74}^u \kappa_{41}^u.
\end{equation}
The node-to-node hidden current from node 1 to 9 is calculated by Eq.~(\ref{eq.3.flow calculation node-node}), and source-to-sink hidden current is calculated by Eq.~(\ref{eq.3.flow calculation source-sink}).

Table~\ref{tab.3.flow tracing result proportion sharing} illustrates the results of flow tracing using the downstream flow tracing method and the upstream flow tracing method.
The numbers in the following table indicate source-to-sink hidden currents.
As we can see, the two methods  imply the same results.

\begin{table}[htb]
\centering
\caption[Flow tracing results]{Flow tracing in the DC network shown in Fig.~\ref{fig.3.DC for simulink}, where nodes 1 and 2 are source nodes, and nodes 9 and 10 are sink nodes. Numbers in the table shows source-to-sink hidden flows.}
\label{tab.3.flow tracing result proportion sharing}
\begin{tabular}{|c|c|c||c|c|c|}
\hline
\multicolumn{3}{|c||}{Downstream} & \multicolumn{3}{c|}{Upstream} \\ \hline
     Node  &    9   &    10   &   Node    &    9   &    10  \\ \hline
      1 &   2.35    &   0.65    &    1   &   2.35    &  0.65     \\ \hline
      2 &    2.16   &   2.84    &    2   &  2.16     &  2.84     \\ \hline
\end{tabular}
\end{table}
\subsection{Using the Downstream Extended Incidence Matrix}
From the MATLAB simulation results of the DC network, the downstream extended incidence matrix, $\mf{K}$, is 
\begin{center}
$\mf{K}=\begin{bmatrix}
1         & -0.0378 & 0  & 0  & 0  & 0  & 0    & 0  & 0 & 0 \\
0         & 1         & 0  & 0  & 0  & 0  & 0    & 0  & 0 & 0 \\
-0.4571 & 0         & 1  & 0  & 0  & 0  & 0    & 0  & 0 & 0 \\
-0.5429 & -0.6722 & 0  & 1  & 0  & 0  & 0    & 0  & 0 & 0 \\
0         & -0.2900 & 0  & 0  & 1  & 0  & 0    & 0  & 0 & 0 \\
0         & 0         & -1 & 0  & 0  & 1  & 0    & 0  & 0 & 0 \\
0         & 0         & 0  & -1 & 0  & 0  & 1    & 0  & 0 & 0 \\
0         & 0         & 0  & 0  & -1 & 0  & 0    & 1  & 0 & 0 \\
0         & 0         & 0  & 0  & 0  & -1 & -0.6000 & 0  & 1 & 0 \\
0         & 0         & 0  & 0  & 0  & 0  & -0.4000 & -1 & 0 & 1
\end{bmatrix},$
\end{center}
and the downstream contribution matrix, $\mf{C}$, is
\begin{center}
$\mf{C}=\begin{bmatrix}
1        & 0.0378 & 0 & 0   & 0 & 0 & 0   & 0 & 0 & 0 \\
0        & 1        & 0 & 0   & 0 & 0 & 0   & 0 & 0 & 0 \\
0.4571 & 0.0173 & 1 & 0   & 0 & 0 & 0   & 0 & 0 & 0 \\
0.5429 & 0.6927 & 0 & 1   & 0 & 0 & 0   & 0 & 0 & 0 \\
0        & 0.2900 & 0 & 0   & 1 & 0 & 0   & 0 & 0 & 0 \\
0.4571 & 0.0173 & 1 & 0   & 0 & 1 & 0   & 0 & 0 & 0 \\
0.5429 & 0.6927 & 0 & 1   & 0 & 0 & 1   & 0 & 0 & 0 \\
0        & 0.2900 & 0 & 0   & 1 & 0 & 0   & 1 & 0 & 0 \\
0.7828 & 0.4329 & 1 & 0.6000 & 0 & 1 & 0.6000 & 0 & 1 & 0 \\
0.2172 & 0.5671 & 0 & 0.4000 & 1 & 0 & 0.4000 & 1 & 0 & 1
\end{bmatrix}.$
\end{center}
We also obtain, from the experiments, that $f_1^{in}=3.1891$, $f_2^{in}=5$, $\iota_1^s=0.9407$, $\iota_2^s=1$, $\iota_9^t=1$ and $\iota_{10}^t=1$.
Thus, we calculate $f_{sj\rightarrow ti}$ for $j=1,~2$ and $i=9,~10$ by $f_{s1\rightarrow t9}=\iota_9^t\cdot C_{91} f_1^{in} \cdot \iota_1^s=2.35$, 
$f_{s2\rightarrow t9}=\iota_9^t\cdot C_{92} f_2^{in} \cdot \iota_2^s=2.16$, 
$f_{s1\rightarrow t10}=\iota_{10}^t\cdot C_{10~1} f_1^{in} \cdot \iota_1^s=0.65$, and 
$f_{s2\rightarrow t10}=\iota_{10}^t\cdot C_{10~2} f_2^{in} \cdot \iota_2^s=2.84$.
We note that all these numbers coincide with that in Tab.~\ref{tab.3.flow tracing result proportion sharing}. 

\subsection{Using the Upstream Extended Incidence Matrix}
Define the \textit{upstream extended incidence matrix}, $\mf{K'}$, by
\begin{equation}
\label{eq.3.upstream extended incidence matrix}
K'_{ij}=
\begin{cases}
-f_{ij}^{out}/f_j^{in} & ~\text{if}~i\neq j,~\text{and}~f_{ij}>0,\\
1 & ~\text{if}~i=j,\\
0 & ~\text{else}.
\end{cases}
\end{equation}
We know $f_i^{out}=\sum_{j=1}^{N} f_{ij}^{out}+f_i^t$, implying,
$
f_i^{out}-\sum_{j=1}^{N} f_{ij}^{out}/f_j^{in} \cdot f_j^{in}=f_{i}^t.
$
Since $f_i^{out}= f_i^{in}$, we have
\begin{equation}
\label{eq.3.upstream equaiton}
f_i^{in}-\sum_{j=1}^{N} f_{ij}^{out}/f_j^{in}\cdot f_j^{in}=f_{i}^t.
\end{equation}
Equations~(\ref{eq.3.upstream extended incidence matrix}) and (\ref{eq.3.upstream equaiton}) imply
\begin{equation}
\label{eq.3.upstream matrix equation}
\mf{K' F^{in}=F^t},
\end{equation}
where $\mf{F^{in}}=[f_{1}^{in},~f_{2}^{in},\cdots, f_{N}^{in}]^T$ and $\mf{F^t}=[f_1^t,~f_2^t,\cdots,f_N^t]^T$.
From $\mf{F^{in}=K'^{-1}F^t}$, we have 
\begin{equation}
\label{eq.3.upstream matrix coefficient}
\begin{split}
f_{i}^{in}&=\sum_{j=1}^N\left[\mf{K'^{-1}}\right]_{ij} f_j^t\\
&=\sum_{j=1}^N\left[\mf{K'^{-1}}\right]_{ij}f_j^{out} \cdot\iota_j^t.
\end{split}
\end{equation}
Let $\mf{C'=K'^{-1}}$ be the \textit{upstream contribution matrix} whose element, $C_{ij}=\left[\mf{K'^{-1}}\right]_{ij}$, is a \textit{upstream contribution factor} indicating how much proportion of the total outflow at node $j$ is coming from node $i$, i.e., $f_{i\rightarrow j}=C'_{ij} f_j^{out}$.
Then, $f_{si\rightarrow tj}=\iota_i^s \cdot C'_{ij}f_j^{out}\cdot \iota_j^t$. 

The upstream extended incidence matrix, $\mf{K'}$, of the DC network is 
\begin{center}
$\mf{K}=\begin{bmatrix}
1         & 0 & -1 & -0.3400 & 0  & 0  & 0  & 0  & 0         & 0         \\
-0.0593 & 1 & 0  & -0.6600 & -1 & 0  & 0  & 0  & 0         & 0         \\
0         & 0 & 1  & 0         & 0  & -1 & 0  & 0  & 0         & 0         \\
0         & 0 & 0  & 1         & 0  & 0  & -1 & 0  & 0         & 0         \\
0         & 0 & 0  & 0         & 1  & 0  & 0  & -1 & 0         & 0         \\
0         & 0 & 0  & 0         & 0  & 1  & 0  & 0  & -0.3230 & 0         \\
0         & 0 & 0  & 0         & 0  & 0  & 1  & 0  & -0.6770 & -0.5842 \\
0         & 0 & 0  & 0         & 0  & 0  & 0  & 1  & 0         & -0.4158 \\
0         & 0 & 0  & 0         & 0  & 0  & 0  & 0  & 1         & 0         \\
0         & 0 & 0  & 0         & 0  & 0  & 0  & 0  & 0         & 1        
\end{bmatrix},$
\end{center}
and the upstream contribution matrix, $\mf{C'}$, is
\begin{center}
$\mf{C'}=\small{\begin{bmatrix}
1        & 0 & 1        & 0.3400 & 0 & 1        & 0.3400 & 0 & 0.5532 & 0.1986 \\
0.0593 & 1 & 0.0593 & 0.6802 & 1 & 0.0593 & 0.6802 & 1 & 0.4796 & 0.8132 \\
0        & 0 & 1        & 0        & 0 & 1        & 0        & 0 & 0.3230 & 0        \\
0        & 0 & 0        & 1        & 0 & 0        & 1        & 0 & 0.6770 & 0.5842 \\
0        & 0 & 0        & 0        & 1 & 0        & 0        & 1 & 0        & 0.4158 \\
0        & 0 & 0        & 0        & 0 & 1        & 0        & 0 & 0.3230 & 0        \\
0        & 0 & 0        & 0        & 0 & 0        & 1        & 0 & 0.6770 & 0.5842 \\
0        & 0 & 0        & 0        & 0 & 0        & 0        & 1 & 0        & 0.4158 \\
0        & 0 & 0        & 0        & 0 & 0        & 0        & 0 & 1        & 0        \\
0        & 0 & 0        & 0        & 0 & 0        & 0        & 0 & 0        & 1  
\end{bmatrix}}.$
\end{center}
We also obtain $f_9^{out}=4.5132$, $f_{10}^{out}=3.4868$, $\iota_1^s=0.9407$, $\iota_2^s=1$, $\iota_9^t=1$ and $\iota_{10}^t=1$.
Then, 
$f_{s1\rightarrow t9}=\iota_1^s\cdot C'_{19} f_9^{out} \cdot \iota_9^t=2.35$, 
$f_{s2\rightarrow t9}=\iota_2^s\cdot C'_{29} f_{9}^{out} \cdot \iota_{9}^t=2.16$,
$f_{s1\rightarrow t10}=\iota_1^s\cdot C'_{1~10} f_{10}^{out} \cdot \iota_{10}^t=0.65$, and 
$f_{s2\rightarrow t10}=\iota_2^s\cdot C'_{2~10} f_{10}^{out} \cdot \iota_{10}^t=2.84$.
The results are the same as that in Tab.~\ref{tab.3.flow tracing result proportion sharing}.
\end{document}